\documentclass[11pt]{article}

\usepackage{graphicx}  
\usepackage{amsmath}
\usepackage{times}
\usepackage{authblk}
\usepackage{geometry}
\title{Critical Josephson current in the dynamical Coulomb blockade regime}

\author[1]{Berthold J\"ack \thanks{berthold.jaeck@hotmail.de}}
\author[1]{Matthias Eltschka}
\author[1]{Maximilian Assig}
\author[1]{Markus Etzkorn}
\author[1]{Christian R. Ast}
\author[1,2]{Klaus Kern}
\affil[1]{Max-Planck-Institut f\"ur Festk\"orperforschung, 70569 Stuttgart, Germany}
\affil[2]{Institut de Physique de la Mati{\`e}re Condens{\'e}e, Ecole Polytechnique F{\'e}d{\'e}rale de Lausanne, 1015 Lausanne, Switzerland}

\begin{document}
\maketitle

\begin{abstract}
Superconductivity is commonly described as a macroscopic quantum phenomenon. However, it arises from microscopic mechanisms occurring at the nanometer scale as illustrated, for example, by the non-trivial pairing in unconventional superconductors. More recently, also local interactions with superconductors in the context of Majorana fermions became of interest. A very direct way to study the atomic scale properties of superconductors is given by the combination of the Josephson effect with scanning tunneling microscopy (STM), also referred to as JSTM. Here, the critical Josephson current serves as a direct local probe of the superconducting ground state and may reveal valuable information that is often inaccessible when studying quasi-particle excitation spectra. We show that we can extract local values of the critical Josephson current from JSTM measurements in the dynamical Coulomb blockade regime. Furthermore, we experimentally determine the regime of sequential Cooper pair tunneling, which is in accordance to theoretical predictions. Our study presents new insights on the tunneling mechanisms in Josephson junctions and lays the basis for the implementation of JSTM as a versatile probe for superconductivity.
\end{abstract}

The DC Josephson effect describes the tunneling of Cooper pairs between two superconducting electrodes, which manifests itself as a finite tunneling current at zero voltage \cite{Josephson_1962}. The maximum amplitude of this current, the critical Josephson current $I_{0}$, directly depends on the normal state conductance $G_{\text{N}}$ of the tunnel contact and on the superconducting order parameters $\Delta$ of the electrodes \cite{Ambegaokar_1963}. Due to the direct dependence on $\Delta$, the critical Josephson current represents excellent means to {\em directly probe} the superconducting ground state of a sample under investigation and may reveal still unknown and valuable information that is inaccessible when studying quasi-particle excitation spectra. It can shed light on the superconducting ground state of unconventional superconductors, such as cuprates, and on the properties of novel superconductors like the single-layer iron selenide on STO \cite{Smakov_2001, Kun_2012, Xue_2015}. For the latter, recent scanning tunneling microscopy (STM) experiments on the quasi-particle characteristics indicate an s-wave pairing symmetry of the superconducting state \cite{Feng_2015}. Moreover, the critical Josephson current may also be used to probe spatial variations in $\Delta$ providing insight on the local interaction of superconductors with magnetic impurities \cite{Yazdani_1997, Flatte_1997,salkola_spectral_1997}, a topic that recently gained significant scientific interest due to the experimental observation of Majorana fermions in ferromagnetic iron chains on lead \cite{Yazdani_2014}. Here, a local probe of superconductivity can test for the p-wave type of superconductivity on the iron chain, which is associated with the occurrence of Majorana bound states. Therefore, the combination of the Josephson effect with the atomic scale resolution of low temperature STM, also referred to as Josephson STM (JSTM) \cite{Smakov_2001}, holds promising potential as probe for superconductivity and related phenomena that are in scope of the scientific community. In first JSTM experiments the tunneling of Cooper pairs through the atomic scale tunnel junction was demonstrated \cite{Naaman_2001, Vieira_2006, Marcus_2008, Kimura_2008, Wellstood_2015} and also the spatial mapping of this current was realized successfully \cite{Proslier_2006}. Determining quantitative values of $I_0$ from JSTM experiments, however, has not been achieved so far, although this capability is of fundamental importance for the concept of JSTM \cite{Smakov_2001}. One possibility to extract $I_0$ from experimental data is given by the Ivanchenko and Zil'berman model \cite{zilberman_1969,Steinbach_2001}, if the capacitance can be neglected. However, in a typical STM geometry, the junction capacitance cannot be neglected \cite{Jaeck_2015}. Under these conditions, the so-called $P(E)$-theory \cite{Devoret_1990, Averin_1990} has to be used to describe the tunneling current. This has been demonstrated before both in the context of single-particle tunneling \cite{Brun_2012, Roditchev_2013} as well as sequential Cooper pair tunneling \cite{Jaeck_2015}.

In the following, we demonstrate that the local value of the critical Josephson current, as extracted from the fits of the $P(E)$-theory to the experimental data from JSTM experiments, corresponds to the value from the Ambegaokar-Baratoff (AB) formula. Further, we experimentally observe a regime in which the phase tunneling starts to dominate the sequential Cooper pair tunneling and where $P(E)$-theory breaks down. In this way, we experimentally determine the range of sequential Cooper pair tunneling, which is in agreement to theoretical predictions. In the context of JSTM, this result also allows us to establish an optimal parameter range, in which JSTM experiments can be performed.
\\
\\{\bf Fundamental Considerations and Theory}\\ The current-voltage-characteristics of a Josephson junction generally depends on a number of different parameters, which requires a careful choice of the theoretical model \cite{zilberman_1969, Averin_1990, Devoret_1990}. To do this, we compare the different energy scales of all involved physical phenomena. These are the Josephson coupling energy $E_{\text{J}}=\hbar I_{\text{0}}/(2e)$ ($\hbar$ is the reduced Planck constant $\hbar=h/(2\pi)$ and $e$ is the elementary charge), the Coulomb charging energy of the tunnel contact $E_{\text{C}}=2e^2/C_{\text{J}}$, where $C_{\text{J}}$ is the junction capacitance, as well as the thermal energy $E_{\text{T}}=k_{\text{B}}T$, where $T$ is the temperature and $k_{\text{B}}$ is the Boltzmann constant. The Josephson energy $E_{\text{J}}$ in our case is on the order of 10\,$\mu$eV, in the tunneling regime where $G_{\text{N}}\ll G_{\text{0}}$ ($G_{\text{N}}$ is the normal state conductance and $G_{\text{0}}=2e^2/h$ denotes the quantum of conductance). The Coulomb charging energy $E_{\text{C}}$ is on the order of 100\,$\mu$eV assuming a typical STM junction capacitance $C_{\text{J}}$ of a few femtofarad. At an effective temperature of 40\,mK, the thermal energy $E_{\text{T}}$ is 3.45\,$\mu$eV \cite{Assig_2013}.

\begin{figure}[h!]
\centering
\includegraphics[width=.5\columnwidth]{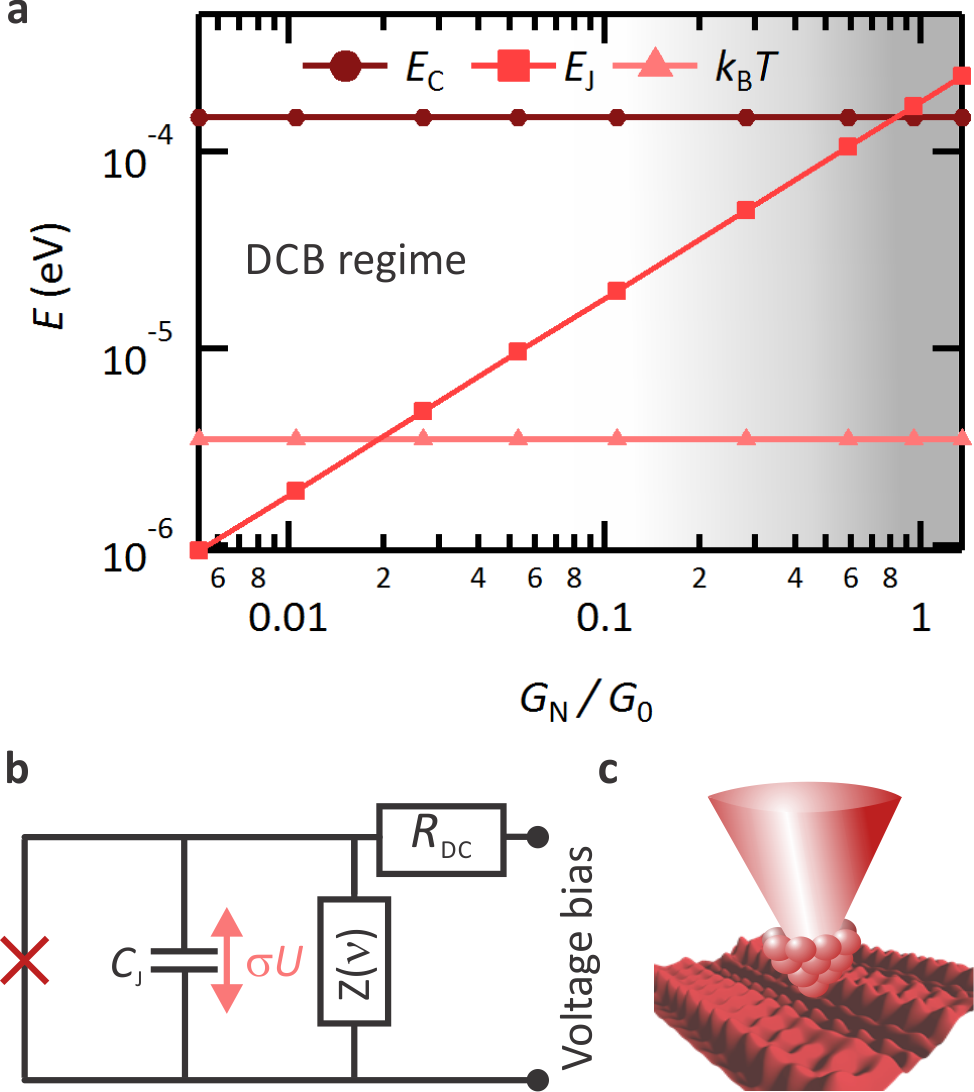}
\caption{\textbf{Properties of a superconducting tunnel junction in an STM.} \textbf{a,}  Coulomb charging energy $E_{\text{C}}$, Josephson coupling energy $E_{\text{J}}$ and thermal energy $E_{\text{T}}$ at $T_\text{eff}\leq40\,\text{mK}$ as a function of the normalized tunnel conductance $G_{\text{N}}/G_{\text{0}}$. $E_{\text{C}}$ was calculated using the average value of the fitted capacitance. \textbf{b,} Simplified circuit diagram of the experimental setup: The red cross represents the Josephson junction, $C_{\text{J}}$ the junction capacitance, $\sigma U$ the voltage noise, $Z(\nu)$ the environmental impedance and $R_{\text{DC}}$ the voltage drop on the circuit. \textbf{c,} The surface topography shows the $(5\times1)$ reconstructed V(100) surface measured at a tunnel setpoint of $V'=2\,\text{mV}$ and $I=5\,\text{nA}$. Above the sample, an artistic view of the vanadium STM tip is shown.}
\label{fig_1}
\end{figure}

Figure\,\ref{fig_1}a compares these energy scales in our experiment for different values of $G_{\text{N}}$. We find that in the tunnel regime ($G_{\text{N}}\ll G_{\text{0}}$), the energy scales order in the following way: $E_{\text{T}}\ll E_{\text{J}} \ll E_{\text{C}}$. In particular, this means that the condition $E_{\text{T}}\le E_{\text{J}}$ for JSTM to work best is fulfilled for most of the tunnel conductance range \cite{zilberman_1969,Smakov_2001}. In addition, in the limit $E_{\text{J}}\ll E_{\text{C}}$, the tunneling current is created by the sequential tunneling of Cooper pairs, also referred to as the dynamical Coulomb blockade (DCB) regime. In this regime, the Cooper pairs tunnel inelastically releasing energy quanta $h\nu$ proportional to the junction bias voltage $V_{\text{J}}=h\nu/(2e)$ into the environment. The emitted photon spectrum has recently been studied in more detail \cite{hofheinz_2011}, also in the context of non-linear quantum dynamics \cite{gramich_2013}. The sequential Cooper pair tunneling characteristics can be modeled by the $P(E)$-theory \cite{Averin_1990, Devoret_1990}, which treats the Josephson coupling energy $E_{\text{J}}$ as a perturbation to the Coulomb energy $E_{\text{C}}$. This theory facilitates the determination of an experimental Josephson coupling energy $E_{\text{J}}$, which can be directly converted to the Josephson critical current $I_{\text{0}}=(2e/\hbar)E_{\text{J}}$ -- giving access to $\Delta$ \cite{Averin_1990, Devoret_1990}. However, it is \textit{a priori} not clear that the experimental values of $I_{\text{0}}$ found in the DCB regime will correspond to the actual AB critical current, that has been evaluated for the phase-tunneling regime \cite{Ambegaokar_1963,Joyez_2013}. Moreover, when the Josephson coupling energy $E_{\text{J}}$ becomes comparable to $E_{\text{C}}$ -- in our case when $G_{\text{N}}\approx G_{\text{0}}$ (see Fig.\,\ref{fig_1}a) -- the Josephson junction enters a regime, where phase tunneling becomes more and more dominant. Therefore, $P(E)$-theory, describing sequential Cooper pair tunneling, should fail to describe $I(V)$-characteristics measured in this regime, which remains an unresolved question until now.

The perturbative approach of $P(E)$-theory applies Fermi's golden rule to calculate the tunneling current \cite{Grabert_1994}:
\begin{equation}
I(V)=\frac{\pi e}{\hbar}E_{\text{J}}^{2}\left[P(2eV)-P(-2eV)\right],
\label{eq:poe}
\end{equation}
where $P(E)$ is the spectral probability for a tunneling Cooper pair to emit ($E>0$) or absorb $(E<0)$ a photon to or from the electromagnetic environment, whose circuit diagram is shown in Fig.\,\ref{fig_1}b. The probability distribution $P(E)$ is only determined by the electromagnetic environment $Z(\nu)$ of the junction and independent of the normal state conductance $G_{\text{N}}$ \cite{Devoret_1990, Averin_1990}. The Josephson effect enters \textit{only} through the scaling factor $E_{\text{J}}^2$, which is particularly advantageous for the following data analysis: $I(V)$-curves measured at different values of $G_{\text{N}}$ can be modeled by the \textit{same} $P(E)$-function scaled by $E_{\text{J}}^2$. We will use this property later to mark the range of validity of $P(E)$-theory. And, $E_{\text{J}}$ is independent of $Z(\nu)$, for which reason its value can be unambiguously determined with high precision. The probability $P(E)$ in Eq.\ \ref{eq:poe}, whose energy integral normalizes to one, is a convolution of two independent energy exchange probabilities $P_{\text{Z}}(E)$ and $P_{\text{C}}(E)$ \cite{Grabert_1994, Ingold_1991, Ast_2015}. The probability $P_{\text{Z}}(E)$ describes the energy exchange with the immediate environment, which is characterized by a complex, frequency dependent impedance $Z_{\text{T}}(\nu)$. In the STM, it consists of the junction capacitance $C_{\text{J}}$ as well as the tip, which acts as a $\lambda/4$-monopole antenna. It can be modeled effectively by a modified open-ended transmission line impedance \cite{Jaeck_2015} (also see methods section for more details). Moreover, phase diffusion effects due to finite temperature in the resistive leads are incorporated in $P_{\text{Z}}(E)$ through an ohmic contribution $R=Z_{\text{T}}(0)$ at zero frequency. The second distribution $P_{\text{C}}(E)$ accounts for an experimentally observed broadening of the Cooper pair current spectrum. A likely source of this broadening are thermal charge fluctuations in the junction electrodes resulting in thermal voltage fluctuations $\sigma U$ across the junction capacitance (see Fig.\,\ref{fig_1}b). We estimate the corresponding $P_{\text{C}}(E)$ function to be of Gaussian shape with a standard deviation of $\sigma=\sqrt{2E_{\text{C}}k_{\text{B}}T}$ \cite{Ast_2015}. We will show that the contribution from the thermal voltage fluctuations is essential for modeling the $I(V)$-curves. 

\begin{figure}[h!]
\centering
\includegraphics[width=1\columnwidth]{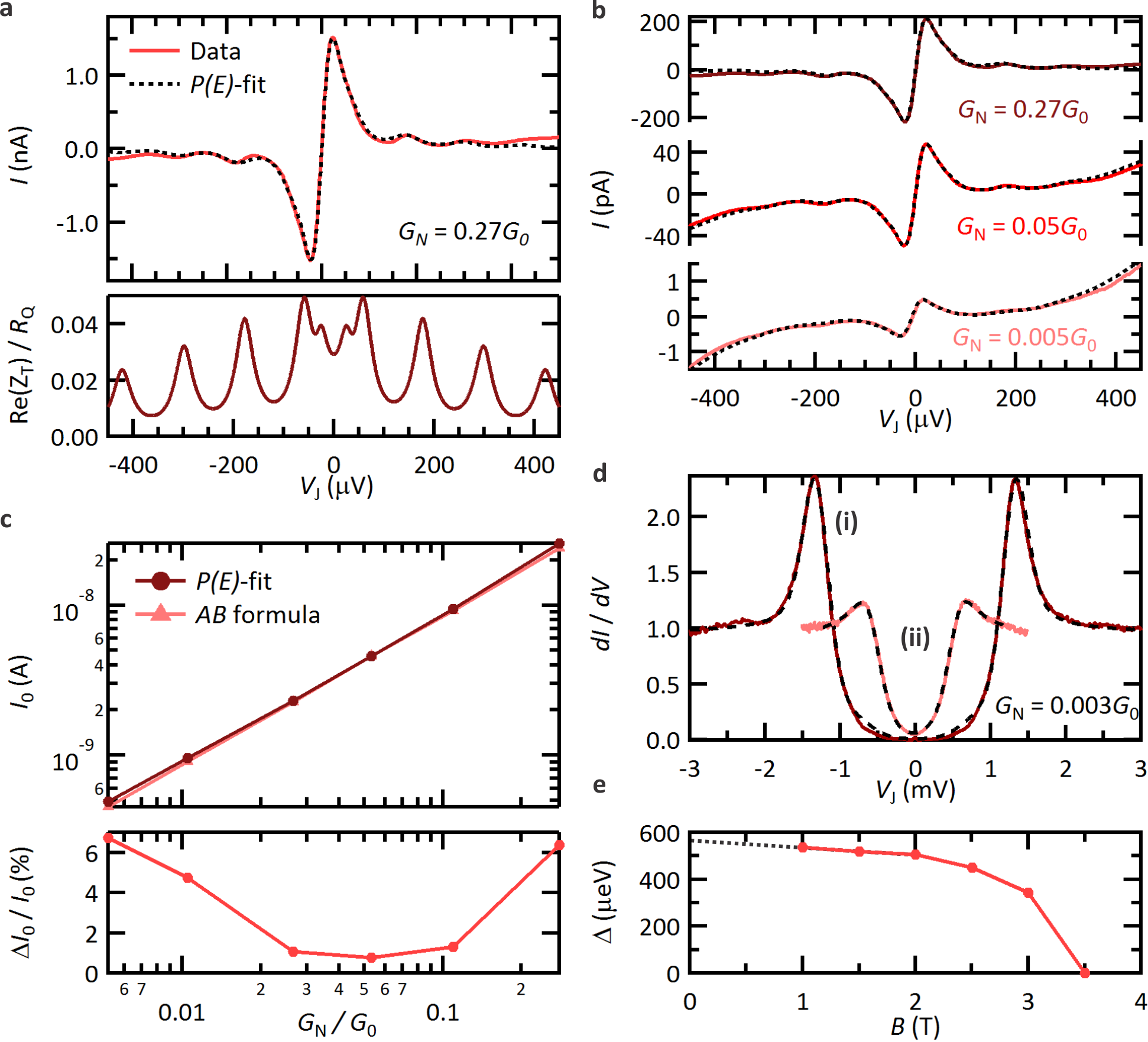}
\caption{\textbf{Experimental values of the Josephson critical current.} \textbf{a,}  Typical $I(V)$-curve and its $P(E)$-fit as well as the corresponding real part of $Z_{\text{T}}(\nu)$ (cf. Ref.\cite{Jaeck_2015}). \textbf{b,} $I(V)$-curves (solid lines) measured at indicated tunnel conductance values $G_{\text{N}}$ and the corresponding $P(E)$-fits (dashed lines). \textbf{c,} Experimentally determined ($P(E)$-fit) and calculated (AB formula) values of $I_{\text{0}}$ as well as their relative deviation $\Delta I_{\text{0}}/I_{\text{0}}$ plotted as a function of the normalized conductance $G_{\text{N}}/G_{\text{0}}$. \textbf{d,}(i) Normalized $dI/dV$-spectrum (solid line) and the corresponding Dynes-fit (dashed line) for superconducting tip and sample. (ii) Normalized $dI/dV$-spectrum (solid line) and the corresponding Maki-fit (dashed line) of the superconducting tip measured at a magnetic field of $B=1\,\text{T}$. \textbf{e,} Values of the STM tip's order parameter $\Delta_2$, as extracted from the Maki fits as a function of the applied magnetic field $B$. The extrapolation to zero field is indicated by the dashed line.}
\label{fig_2}
\end{figure}

In the following we will present experimental results on the Josephson Effect in the microscopic tunnel junction of an STM operated at an effective electronic temperature of $T\leq40\,\text{mK}$ \cite{Assig_2013}. For this study, the Josephson junction consists of a poly-crystalline vanadium (V) STM tip and an atomically clean V(100) single crystal as the STM sample placed in an UHV environment, as is shown in Fig.\,\ref{fig_1}c (also see methods section for more details).
\\
\\
{\bf Experimental results and discussion}\\ A typical $I(V)$-curve measured at a conductance of $G_{\text{N}}=0.27\,G_{\text{0}}$ is shown in Fig.\ \ref{fig_2}(a). The $I(V)$-curve features a dominant supercurrent peak near zero voltage and well-defined spectral resonances at higher voltages, which originate from the interaction of the junction with the tip-assembly impedance \cite{Devoret_1990, Holst_1994, Jaeck_2015}. Moreover, in comparison with previous studies, e.\ g.\ References \cite{Holst_1994, Steinbach_2001}, all current features exhibit a rather broad contour, which can be attributed to the intrinsically low quality factor of antennas as well as the impact of the voltage fluctuations $\sigma U$. The challenge in fitting an $I(V)$-curve using $P(E)$-theory lies in the rather complex interplay of the different fitting parameters, which require a more detailed consideration. The resistive junction leads are transmission lines, for which reason we can set the dissipative impedance at zero frequency $Z_{\text{T}}(0)$ to the input impedance of a transmission line, $R_{\text{env}}=377\,\Omega$ \cite{Joyez_1999}. Further, we use an effective electronic temperature of $T_{\text{eff}}=40\,\text{mK}$ that we determined independently \cite{Assig_2013}. To account for in-gap quasiparticle contributions to the tunnel current, we also add a cubic background to the current in Eq.\,\ref{eq:poe}.

Incorporating these parameters, we can fit the experimental $I(V)$-curve as shown in Fig.\,\ref{fig_2}a. The fit reproduces both the supercurrent peak as well as the spectral resonances with high accuracy and we can extract a Josephson coupling energy of $E_{\text{J}}=52.69\pm0.53\,\mu\text{eV}$. The environmental impedance $Z_{\text{T}}(\nu)$, whose real part is displayed in Fig.\,\ref{fig_2}a, shows its base resonance frequency at $\nu_{0}=31.34\pm0.04\,\text{GHz}$ and a corresponding damping factor $\alpha=0.52\pm0.01$ (cf.\ \cite{Jaeck_2015}). For the junction capacitance, we find a typical value of $C_{\text{J}}=2.04\pm 0.07\,\text{fF}$. We conclude that the $I(V)$-curves from our small capacitance tunnel junction showing the characteristics of Cooper pair tunneling can be described by $P(E)$-theory with high accuracy and reasonable parameters, that are independently reproducible. Moreover, we are able to unambiguously determine an experimental value of the Josephson coupling energy $E_{\text{J}}$ in a particular junction.

We have repeated the same analysis for several Cooper pair tunneling characteristics over a large range of the normal state tunneling conductance $0.0052\,G_{\text{0}}\leq G_{\text{N}}\leq1.35\,G_{\text{0}}$. The measured $I(V)$-curves were fitted with $P(E)$-theory in the same fashion as before, of which three examples are shown in Fig.\,\ref{fig_2}b. For all values of $G_{\text{N}}$, $P(E)$-theory describes the tunneling current with high accuracy. From the fitted Josephson coupling energy $E_{\text{J}}$, we can directly calculate an experimental critical current $I_{\text{0}}=2e/\hbar E_{\text{J}}$. Its dependence on the normal state tunneling conductance $G_{\text{N}}$ is displayed in Fig.\,\ref{fig_2}c. We find that $I_{\text{0}}$ linearly depends on $G_{\text{N}}$ over almost two orders of magnitude for $G_{\text{N}}\leq 0.27G_{\text{0}}$. As underlined above, the $P(E)$-distribution is independent of $G_{\text{N}}$ (cf.\ Eq.\ \ref{eq:poe}). Hence, we can assign this linear increase of $I_{\text{0}}$ entirely to the increase of $G_{\text{N}}$, which is in agreement with the AB formula \cite{Ambegaokar_1963}.

\begin{figure}[h!]
\centering
\includegraphics[width=.5\columnwidth]{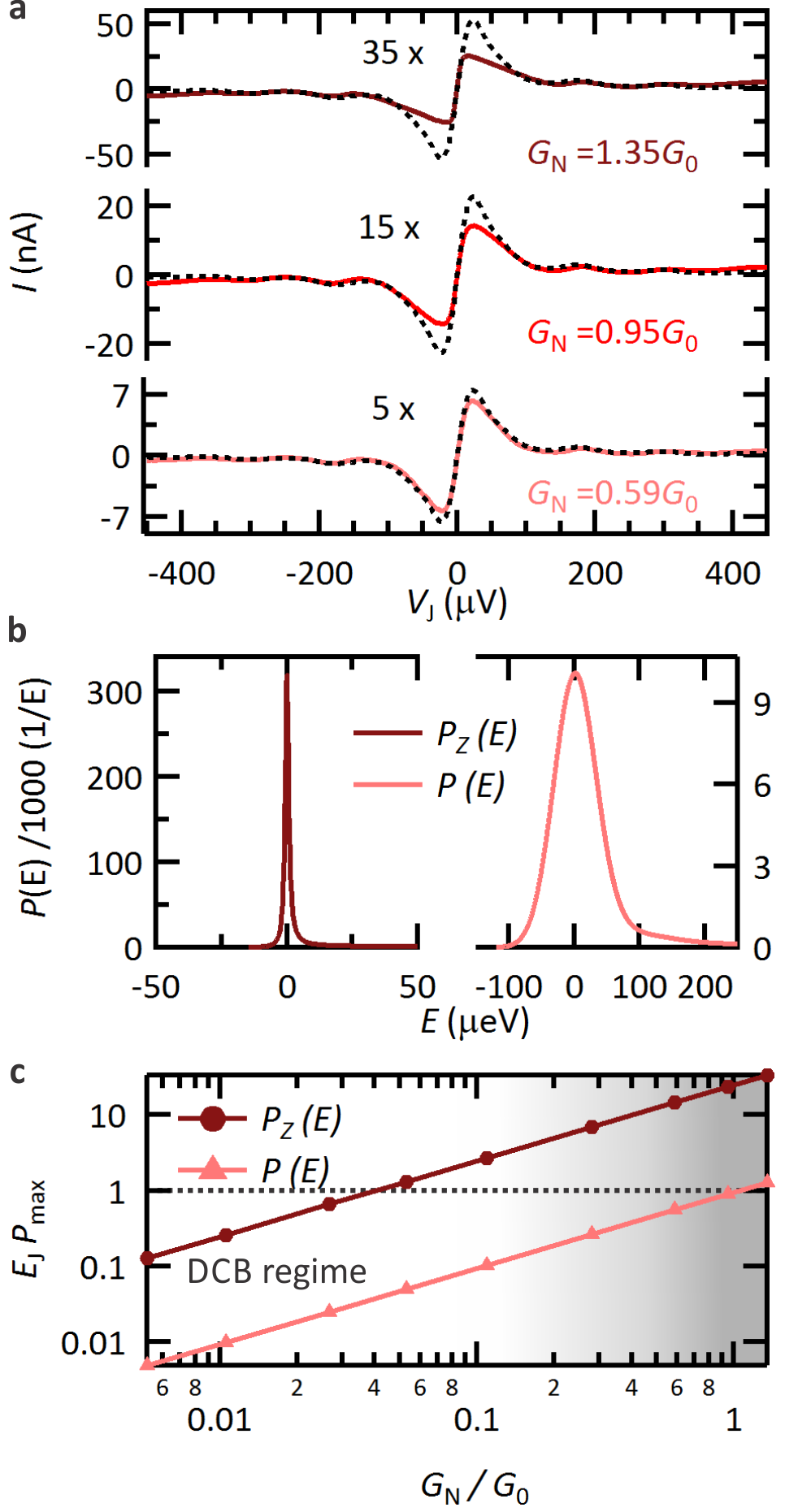}
\caption{\textbf{Regime of sequential Cooper pair tunneling.} \textbf{a,} $I(V)$-curves (solid lines) and up-scaled fits (dashed lines) measured at large values of $G_{\text{N}}/G_{\text{0}}$. The scaling factors with respect to the fit at $G_{\text{N}}=0.27\,G_{\text{0}}$ are indicated. \textbf{b,} Calculated probability distribution $P_{Z}(E)$, only considering the dissipative environment, and the total distribution $P(E)$ that also considers the capacitive voltage noise $\delta U$. Note the different scales in the two panels. \textbf{c,} The products $E_{\text{J}}P_{Z,\text{max}}$ and $E_{\text{J}} P_\text{max}$ plotted as a function of the normalized conductance $G_{\text{N}}/G_{\text{0}}$ indicating the range of validity of $P(E)$-theory.}
\label{fig_3}
\end{figure}

To quantitatively compare the experimentally found values for the critical current with the critical current values calculated from the AB formula, we write the AB formula for two superconductors with unequal order parameters $\Delta_{1,2}$ and $\Delta_1>\Delta_2$ \cite{Ambegaokar_1963}:
\begin{equation}
I_{0}=\Delta_{2}G_N\,K\left(\sqrt{1-\frac{\Delta_{2}^{2}}{\Delta_{1}^{2}}}\right).
\end{equation}
Here, $K$ denotes Jacobi's full elliptic integral of the first kind. We can independently determine the sample gap $\Delta_1$ and the tip gap $\Delta_2$, by measuring the quasi-particle excitation spectra shown in Fig.\ \ref{fig_3}d at zero external magnetic field (i) and at 1\,Tesla (ii). The sample becomes normal conducting at $B_{\text{c,2}}=0.5\,\text{T}$ \cite{Sekula_1972}, but the tip has a much larger critical field due to the confined geometry at the apex \cite{Meservey_1970}. We, therefore, extract the tip gap by using a Maki model fit for higher fields, as shown in Fig.\ \ref{fig_3}d \cite{Eltschka_2014}. Extrapolating to zero field as shown in Fig.\,\ref{fig_2}e, we find a tip gap of $\Delta_2=563\pm20\,\mu$eV \cite{Eltschka_2015}. The sample gap $\Delta_1$, we can extract from a Dynes fit to the zero field spectrum having the value $\Delta_1=|\Delta_1+\Delta_2|-\Delta_2=748\pm23\,\mu$eV, as shown in Fig.\ \ref{fig_2}d \cite{Dynes_1978}. The reduction of the tip gap compared to the bulk value is common in vanadium tips \cite{Eltschka_2014} and may be explained by the influence of vanadium oxide at the tip surface, changes in the phonon dispersion or grain size effects \cite{Sekula_1972, McMillan_1968, Chen_1969, Strongin_1970}. Inserting these values along with $G_{\text{N}}$ into the AB formula, we can plot the corresponding critical currents in Figure\,\ref{fig_2}c as a function of $G_{\text{N}}$. The critical currents from the $P(E)$-fit and the AB formula match within $<7\%$ (cf.\ lower panel in Fig.\,\ref{fig_2}c) over the entire range of conductance. This is a remarkable observation, since the experimental $I_{\text{0}}$ values were determined in the DCB regime, while the AB formula was derived in the phase-tunneling regime. Our findings confirm the established interpretation of the critical current $I_{\text{0}}$ as a coupling strength between the overlapping pair wavefunctions, which is independent of the actual tunneling process \cite{Ambegaokar_1963}.

In the next step, we tested the range of validity of $P(E)$-theory in the limit $E_{\text{J}}\rightarrow E_{\text{C}}$. Here the initial requirement of this perturbative approach $E_{\text{J}}\ll E_{\text{C}}$ is no longer valid so that $P(E)$-theory should break down. However, Ingold \textit{et al.} found that the global condition $E_{\text{J}}\ll E_{\text{C}}$ is superimposed by another condition $E_{\text{J}} P(E)\ll 1$ \cite{Ingold_1992}. This condition means essentially that sequential tunneling holds as long as the tunneling probability is low enough. In order to test this hypothesis, we measured the $I(V)$-curves for values of the normal state tunneling conductance $G_{\text{N}}\geq0.59\,G_{\text{0}}$ of which three examples are shown in Fig.\,\ref{fig_3}a. Using $P(E)$-theory as before, we were unable to properly fit any of these $I(V)$-curves, which is to be expected, since at the measured conductance values, we find $E_{\text{J}}\approx E_{\text{C}}$ (cf.\ Fig.\,\ref{fig_1}a). Nevertheless, we can up-scale a fitted current spectrum from experiments at a lower conductance $G_{\text{N}}=0.27\,G_{\text{0}}$, since the $P(E)$-function is only determined by the environment. The up-scaled $I(V)$-curve still fits the spectral resonances at higher voltages, but largely overestimates the supercurrent peak around zero voltages in all cases with increasing mismatch for higher values of $G_{\text{N}}$, as shown in Fig.\,\ref{fig_3}a, indicating the breakdown of $P(E)$-theory.

To better understand this observation, we investigated the product $E_{\text{J}} P_\text{max}$, where $P_\text{max}$ is the global maximum of $P(E)$. It is found at zero voltage for the probability distribution of the impedance $P_{\text{Z}}(E)$ as well as the total, convoluted probability distribution $P(E)$ (see Fig.\,\ref{fig_3}b) \cite{Ingold_1992}. It can be seen that the broadening of the total $P(E)$ due to the capacitive noise greatly reduces the maximum value of $P(E)$ compared to $P_{\text{Z}}(E)$. The dependence of $E_{\text{J}} P_\text{max}$ on the tunnel conductance $G_{\text{N}}$ is shown in Fig.\,\ref{fig_3}c. For a conductance of $G_{\text{N}}\geq0.59\,G_{\text{0}}$, we find $E_{\text{J}} P_\text{max}\geq 1$ so that the required condition for $P(E)$-theory is ``locally'' violated near zero voltage. This result perfectly explains our observation that $P(E)$-theory fails to describe the supercurrent peak close to zero bias voltage, where $P(E)$ has its maximum and $E_{\text{J}} P(E\approx 0)\approx 1$. Therefore, we observe phase tunneling at low voltages and charge tunneling at higher voltages in the same spectrum and, thus, $P(E)$-theory fails to model the entire $I(V)$-curve. For quantitative agreement with the experimental data in this regime, higher order perturbation theory may have to be taken into account (see e.g.\ \cite{Johansson_2015}).

Moreover, fitting the $P(E)$-function to our data reveals the significance of the thermal voltage fluctuations as a spectral broadening mechanism. While the probability distribution in the convoluted $P(E)$-function is broadened and has some spectral weight at higher voltages, the $P_{\text{Z}}(E)$ distribution -- only containing the interaction with the dissipative environment -- sharply peaks at $V=0$ (see Fig.\,\ref{fig_3}b). For this reason, the required condition $E_{\text{J}}P_{\text{Z, max}}\ll 1$, is violated for almost the entire conductance range as shown in Fig.\,\ref{fig_3}c. This is in agreement with theory, since we operate the junction in a low impedance environment (i.\ e.\ $Z(0)\ll 1/(2\,G_{\text{0}})$). Therefore, thermal voltage fluctuations have to be included \cite{Ast_2015} to correctly describe our data. This reduces the $P_\text{max}$ values and results in an overall consistent picture between experiment and theory as well as the range of validity.
\\
\\
{\bf Conclusions}\\ In summary, we have investigated the $I(V)$ characteristics of a voltage-biased Josephson junction in the DCB regime with an STM at ultra-low temperatures. We found that the experimentally determined values of the critical Josephson current $I_{\text{0}}=2e/\hbar E_{\text{J}}$ are equal to the theoretical values as calculated by using the Ambegaokar-Baratoff formula. The DCB regime, in which STM experiments are commonly operated, can, therefore, be used to directly determine local absolute values of the critical Josephson current. Furthermore, we experimentally determined the range of sequential Cooper pair tunneling in which $P(E)$-theory can be applied and observed indications for the crossover into the phase tunneling regime at $E_{\text{J}}\geq E_{\text{C}}$. Thus, with precise tuning of the involved energy scales ($E_{\text{T}}$, $E_{\text{J}}$, $E_{\text{C}}$), we can operate our STM in the optimal regime. Our study presents new insights on the tunneling mechanisms in ultra-small Josephson junctions, which lays the basis for the successful implementation of JSTM as a probe for novel topological superconductors in the context of Majorana fermions and for the ground state properties of unconventional superconductors.

It is our pleasure to acknowledge fruitful discussions with F.~Portier, J.~Ankerhold, C.~Urbina and G.-L.~Ingold.

\section*{Methods} \label{exp}
\begin{figure}[h!]
\centering
\includegraphics[width=0.33\columnwidth]{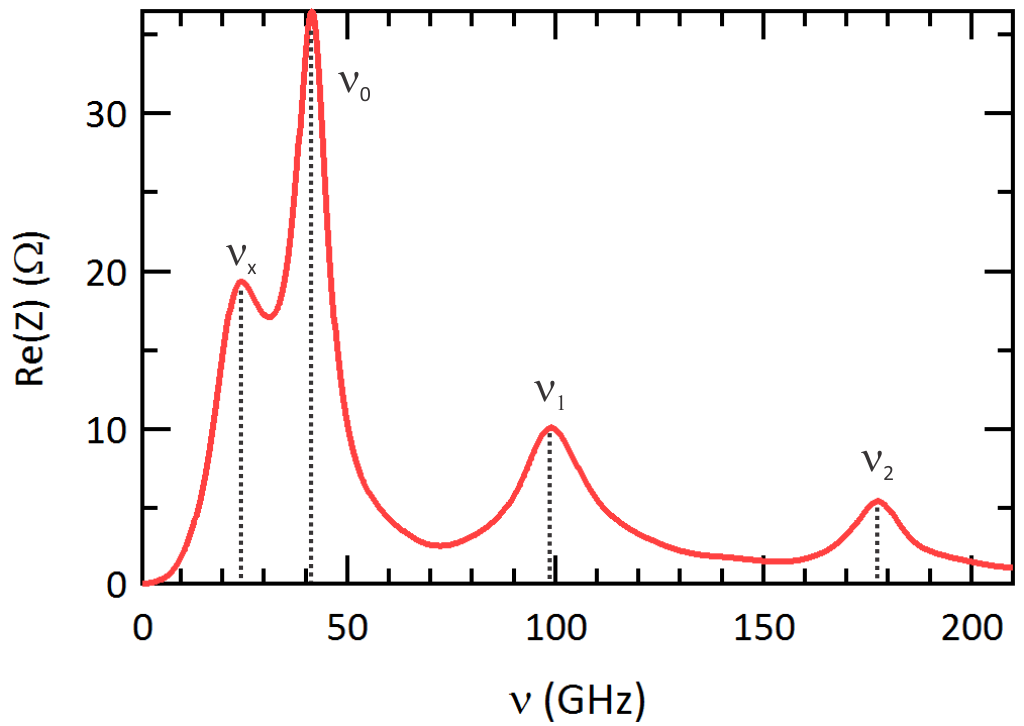}
\caption{Simulated impedance spectrum $Z(\nu)$ as a function of the frequency $\nu$ of the STM tip geometry \cite{Jaeck_2015}. It features the tip resonance modes $\nu_{\text{0,1,2}}$ as well as an additional mode $\nu_{\text{x}}$. The tip length for the calculations was slightly different than the tip length used in the experiment, which is why the resonances are at somewhat different frequencies. Furthermore, the simulation does not consider resistive losses, resulting in a zero DC resistance.}
\label{fig_SI_1}
\end{figure}

{\bf Experimentals}\\
Experiments were performed using an STM with an effective electronic temperature of $T\leq40\,\text{mK}$ \cite{Assig_2013}. For the STM tip, we cut a poly-crystalline vanadium (V) wire of $99.8\,\%$ purity under tension (diameter $d=250\,\mu\text{m}$). The tip was prepared \textit{in situ} by field-emission and voltage-pulses. The sample is a V(100) single crystal \cite{Sekula_1972, Davies_1981}, which has been prepared by cycles of sputtering and annealing to $T=800\,^{\circ}\text{C}$ until it shows an atomically clean ($5\times1$) reconstruction, as shown in Fig.\,\ref{fig_1}b. The normal state tunnel conductance $G_{\text{N}}=I_{\text{T}}/V_{\text{T}}$ is determined by the tunneling current $I_{\text{T}}$ at a bias voltage reference $V_{\text{T}}$, where $eV_{\text{T}}\gg\Delta_1 +\Delta_2$. We correct the voltage axis for voltage drops over an effective circuit resistance $R_{\text{DC}}$, according to $V=V'-I(V')R_{\text{DC}}$. The primed and unprimed voltages denote the applied bias voltage and the junction bias voltage, respectively. The in-line DC resistance of our setup $R_{DC}$ contains experimentally determined contributions from the leads, low pass filters, as well as the input impedance of the current amplifier, which depends on the chosen amplification: $R_{DC}=14\,\text{k}\Omega$ for $G_{N}\leq 0.026G_{0}$, $R_{DC}=4.8\,\text{k}\Omega$ for $0.052G_{0}\leq G_{N}\leq 0.27G_{0}$ and $R_{DC}=3.9\,\text{k}\Omega$ for $G_{N}\geq 0.59G_{0}$. The $dI/dV$-spectra were recorded by standard lock-in techniques applying a modulation frequency of $f=720\,\text{Hz}$ and a modulation amplitude of $V_{\text{mod}}=20\,\mu\text{V}$. Before analyzing the data, the bias axis was corrected for an experimentally determined offset and voltage drops across the measurement circuit.
\\
\\{\bf Environmental impedance}\\ In our STM the surrounding impedance that contributes to the $P(E)$-function is the vacuum as well as the tip acting as a monopole antenna with a corresponding resonance spectrum that depends on the length of the tip \cite{Jaeck_2015}. We approximate the tip assembly impedance $Z(\nu)$ by an infinite transmission line impedance \cite{Grabert_1994, Jaeck_2015} having the analytic expression,
\begin{equation}
Z_{\text{1}}(\nu) = R_{\text{env}} \frac{1+\frac{i}{\alpha}\tan\left(\frac{\pi}{2}\frac{\nu}{\nu_0}\right)}{1+i\alpha\tan\left(\frac{\pi}{2}\frac{\nu}{\nu_0}\right)},
\label{Eq_1}
\end{equation}
where $R_{\text{env}}$ is the effective dc resistance of the environmental impedance, $\alpha$ is an effective damping parameter, and $\nu_{\text{0}}$ is the frequency of the principal resonance. The parameter $R_\text{env}$ is set to the vacuum impedance of 377\,$\Omega$. Yet, finite integral method simulations on the tip holder assembly \cite{Jaeck_2015} show that for specific  geometries of the tip holder, additional resonance features occur in the impedance spectrum of the tip holder assembly, which most likely are hosted by the tip together with the tip holder surface. For instance, Fig.\,\ref{fig_SI_1} shows the real part of the environmental impedance spectrum $Z(\nu)$ as obtained from simulations on the tip holder geometry used in the experiments of this manuscript. It features the typical tip antenna resonance modes $\nu_{\text{0,1,2}}$, at slightly larger frequencies as compared to our experiment, as well as an additional mode $\nu_{\text{x}}$ located at lower frequencies.

While the transmission line impedance in Eq.\,\ref{Eq_1} allows us to model the tip resonances $\nu_{\text{0,1,2}}$, it cannot account for the additional mode. Since the $P(E)$-function represents the energy exchange probability with an environmental impedance, it will be sensitive to modifications of the impedance function. Accordingly, also the measured Cooper pair current will be significantly affected by a modified impedance function. Hence, we include this additional resonance $\nu_{\text{x}}$ into our environmental impedance $Z(\nu)$ in order to correctly fit our experimental data. To this end, we extend the tip impedance $Z_{\text{1}}(\nu)$ as given in Eq.\,\ref{Eq_1} by an additional impedance $Z_{\text{x}}(\nu)$ modeled by \cite{Ingold_1991}:
\begin{equation}
Z_{\text{x}}(\nu) = R_{\text{env}} \frac{1+i\beta^{2}(\nu/\nu_{\text{x}})}{1+i(\nu/\nu_{\text{x}})-\beta^{2}(\nu/\nu_{\text{x}})^{2}}
\label{Eq_2}
\end{equation}  
We can empirically estimate the parameters of this additional impedance $\beta$ and $\nu_{\text{x}}$ from the simulated impedance spectrum in Fig.\,\ref{fig_SI_1}. We set $\nu_{\text{x}}=0.5\nu_{\text{0}}$ and $\beta=2\alpha$, so that we do not introduce any additional fitting parameters to the model. The environmental impedance of the STM tip assembly is the sum of both contributions $Z(\nu)=0.5(Z_{\text{1}}(\nu)+Z_{\text{x}}(\nu))$.
The total impedance $Z_{\text{T}}(\nu)$ as seen by the Josephson junction takes into account the capacitance $C_{\text{J}}$ in the tunnel junction as well:
\begin{equation}
Z_T(\nu)=\frac{1}{i2\pi\nu C_{\text{J}} + Z^{-1}(\nu)}.
\label{eq:zt}
\end{equation}
Hence, the total set of parameters that determine the total environmental impedance are $\alpha$, $\nu_{\text{0}}$ and $C_{\text{J}}$.

\end{document}